\documentclass[10pt]{article}

\usepackage{times}
\usepackage{tikz}
\usepackage{verbatim}
\usepackage{amssymb,amsmath,makeidx,verbatim,graphicx}
\usepackage{float,amsthm,amsfonts,cite,lineno}
\usepackage{graphicx}
\usepackage{epsfig}
\usepackage{subfigure}
\usepackage{float}
\usepackage{url}
\usepackage{amsmath,amsbsy,amsfonts,amssymb}
\usepackage{hhline}
\usepackage{ifthen}
\usepackage{hyperref}
\usepackage{tikz}
\usetikzlibrary{arrows,decorations.pathmorphing,backgrounds,positioning,fit,petri, quotes, shadows, shapes}
\usepackage[absolute,overlay]{textpos}
\everymath{\displaystyle}

\newcommand{\be}{\begin{enumerate}}
\newcommand{\ee}{\end{enumerate}}
\newcommand{\bq}{\begin{eqnarray*}}
\newcommand{\eq}{\end{eqnarray*}}

\catcode `\@=11
\addtocounter{secnumdepth}{1}
\begin{document}
%\begin{linenumbers}
\title{Mathematical Modelling and Analysis of Transmission Dynamics of Lassa fever}
\author{Bakare E.A.$^{\ddagger,\star}$ \footnote{Corresponding author: Email: {\it emmanuel.bakare@fuoye.edu.ng}}, Are E.B.$^{\ddagger,\star}$, Abolarin O.E.$^{\ddagger}$,Osanyinlusi S.A.$^{\dagger}$, Ngwu Benitho $^{\ddagger}$, Ubaka Obiaderi N.  $^{\ddagger}$\\
{\it {\small $^{\star}$Laboratory of Modelling in infectious Diseases and Applied Sciences (LOMIDAS).}}\\
{\it {\small $^{\ddagger}$Department of Mathematics,Federal University Oye Ekiti,}}\\
{\it {\small Ekiti State, Nigeria }}\\
{\it {\small $^{\dagger}$Department of Microbiology,Federal University Oye Ekiti,}}\\
{\it {\small Ekiti State, Nigeria }}\\
}
\date{}
\maketitle

\begin{abstract}
\noindent
Sub-Saharan Africa harbours the majority of the burden of Lassa fever. Clinical diseases, as well as high seroprevalence, has been documented in Nigeria, Sierra Leone, Liberia, Guinea, Ivory Coast, Ghana, Senegal, Upper Volta, Gambia, and Mali. Deaths from Lassa fever occur all year round but naturally peak during the dry season. Annually, the number of people infected is estimated at 100,000 to 300,000, with approximately 5,000 deaths. There have been some few works done on the dynamics of Lassa fever disease transmission but to the best of our knowledge none has been able to capture the seasonal variation of \textit{mastomys} rodents population and its impact on the transmission dynamics. In this work, a periodically-forced seasonal non-autonomous system of a non-linear ordinary differential equation is developed that captures the dynamics of Lassa fever transmission and seasonal variation in the birth of \textit{mastomys} rodents where time was measured in days to capture seasonality. It was shown that the model is epidemiologically meaningful and mathematically well-posed by using the results from the qualitative properties of the solution of the model. A time-dependent basic reproduction number $R_{L}(t)$, is obtained such that its yearly average is written as $\widetilde{R}_{L}<1$, when the disease does not invade the population (means that the number of infected humans always decrease in the following seasons of transmission) and $\widetilde{R}_{L}>1$ when the disease remains constantly and is invading the population and it was detected that $\widetilde{R}_{L} \neq R_{L}$. We also performed some evaluation of the Lassa fever disease intervention strategies using the elasticity of the equilibrial prevalence in order to predict the optimal intervention strategies that can be useful in guiding the local national control program on Lassa fever disease to make a proper decision on the intervention packages. Numerical simulations were carried out to illustrate the analytical results and we found that the numerical simulations of the model showed that possible combined intervention strategies would reduce the spread of the disease. It was established that, to eliminate Lassa fever disease, treatments with Ribavirin must be provided early to reduce mortality and other preventive measures like an educational campaign, community hygiene, Isolation of infected humans, and culling/destruction of rodents must be applied to also reduce the morbidity of the disease. Finally, the obtained results gave a primer framework for planning and designing cost-effective strategies for good interventions in eliminating Lassa fever.
\end{abstract}


\section{Introduction}

Lassa fever (LF) is an acute viral hemorrhagic illness that is common in West Africa. LF is caused by Lassa virus, a single-stranded RNA virus belonging to the family Arenaviridae.  First discovered in 1969 when two missionary nurses died, and named after the Lassa town in Benue State-Nigeria where the ?rst cases occurred, the disease is now endemic in many parts of West African Countries including Nigeria, Sierra Leone, Liberia, and Guinea. Infections with Lassa virus is generally estimated to range from 100,000 to 300,000 with approximately 5,000 deaths each year. The ''multimammate rat" (mastomys natalensis) is regarded as the major reservoir host for the virus in West Africa. This rat specie which is widely distributed throughout the region can shed the virus in its feaces or urine. Humans become infected during direct contact with the rodent reservoir, consumption of food contaminated with rodent feaces and urine or during direct contact with bodily fluids of infected humans.  Signs and symptoms of Lassa fever typically occur 1-3 weeks after the patient comes in contact with the virus. For the majority of Lassa fever virus infections (approximately 80\%), symptoms are mild and are undiagnosed. Mild symptoms include a slight fever, general weakness, and headache. In 20\% of infected individuals, however, the disease may progress to more serious symptoms including hemorrhage (bleeding in gums, eyes, or nose), respiratory distress, repeated vomiting, facial swelling pain in the chest, back and abdomen and shock, neurological problem have also been described including hearing loss, tremors.  Lassa fever is generally treated with the antiviral drug Ribavirin, which has been very effective when given early in the course of the disease (WHO, 2019; CDC 2019).\\

In Nigeria, sporadic outbreaks of Lassa fever have been documented since 1969. The infection is endemic in several states including Edo, Ebonyi, Onitsha, Jos, Taraba, Nasarawa, Yobe, Rivers and Ondo states. In 2012 for example, 623 suspected cases (108 Laboratory con?rmed), including 70 deaths were recorded from 19 states in Nigeria (Nasir and Sani, 2015; WHO, 2012). A total of 11 con?rmed cases of Lassa were recorded in Nigeria with high prevalence in Oyo State in 2014. Between January 1st and 8th of March 2015, the Nigerian Center for Disease Control (NCDC) reported 21 cases of Lassa fever (4Lab. Con?rmed) and 1death due to Lassa (WHO, 2015; CDC, 2015). Between August 2015 and January 2016, there were 239 suspected cases of LF (44 Lab. Con?rmed), including 82 deaths, across 19 states including; Bauchi, Nassarawa, Niger, Delta, Ekiti, Ondo, Kogi, Ebonyi, Lagos, Osun, FCT, Taraba, Kano, Rivers, Edo, Plateau, Gombe, Oyo States etc. (NCDC, 2016; WHO, 2016). Similarly the year 2016, 2017, 2018 and 2019 were not spared from Lassa fever in Nigeria with outbreaks across several states (statistics can be viewed on the Nigerian Center for Disease Prevention and Control website (NCDC)).\\

The outbreak of Lassa fever is highest in humans during the dry season, following Multimammate rodent reservoir breeding during the rainy season.\\

Studies have shown that seasonal timing of reproduction can affect the dynamics of host-pathogen systems (Bolzoni et al. (2006); Dugar et al. (2004); and White et al. (1996)). Seasonality has to do with the systematic peaks of diseases at a certain time of the year and one of its drivers is the birth rate pulses which will be considered in this work. Understanding how seasonally varying parameters act as a forcing mechanism and investigating their dynamical consequences are part of our interest in this work. In which case, we are interested in how such periodically-forced models permits us to better capture the observed pattern of recurrent epidemics different from the unforced models which predict damped oscillation toward the endemic equilibrium (Keeling and Rohani (2008)).\\

Mathematical models have been used severally to capture the dynamics of physical, chemical, biological, economical and many other complex systems. There are various works done so far with the use of mathematical model applied to epidemiology which includes but are not limited to the following; Gumel (2004); Nakul Chitinis et al. (2008); Bakare and Nwozo (2016); Bakare and Nwozo (2017) and many others. Little attention has been paid to this disease in the past leading to scanty information about its transmission dynamics. Despite this, few studies have attempted to study Lassa virus dynamics using mathematical modeling approach. James et al. (2015) studied the dynamics transmission of Lassa fever using susceptible, infected and removed (SIR) model. The authors obtain the disease-free equilibrium and endemic equilibrium states of the system of the differential equation describing the dynamics of the disease. From the stability analyses of the two equilibrium states, they were able to ascertain that if the death rate of the humans (respectively) reservoirs is greater than the respective birth rate that the disease could be control and eradicated. They further specify, from their analysis, when it's practically impossible to control the disease. From this one may ask what happens to the spread if either of the conditions for disease eradication becomes the case. Ogabi et al. (2012) proposed a SIR model for controlling Lassa fever in the northern part of Edo State by considering two senatorial districts of the state with about two million populations. Using numerical approach they analyzed the relationship between the Susceptible, Infected and Removed classes with three health policies. These health policies which consist of three sets of the parameters representing the birth rate, the natural death rate, the transmission rate, and the rate of recovery were employed by the authors to simulate their results. They were able to show that the reproduction number is affected by these parameters, which indirectly tells that 'the health policies' can control the disease, given the role of the reproduction number in disease dynamics. They were able to show also that the disease can be controlled if the transmission rate becomes less than the recovery rate.\\

Bawa et al. (2014) also did a study of Lassa fever dynamics by subdividing the rodent population into infant and adult classes. From their analyses of the disease-free and endemic equilibria, they established a global stability condition for the control of the disease, which is dependent on the reproduction number as obtained in their work. Mohammed et al. (2015) in their work, they developed a transmission dynamic model for Lassa fever with humans immigration. Model analyses was carried out to calculate the reproduction number and sensitivity analysis of the model was also performed. Their results showed that humans immigration rate is the the most sensitive parameter and then the humans recovery rate is the second most sensitive parameter followed by person-to-person contact rate. It was suggested that control strategies should target humans immigration, effective drugs for treatment and education to reduce person-to-person contact. Andrei et al. (2019), in their paper, suggested that seasonal migratory dynamics of rodents played a key role in regulating the cyclic pattern of Lassa fever epidemics but they had no explicit model. Joachim et al. (2019) also suggested in their work that the use of continuous control or rodent vaccination are the strategies that can lead to Lassa fever elimination. Having gone through several works on Lassa fever disease and its transmission dynamics, we observed that none investigated the effect of periodically-forced per-capita birth rate of \textit{mastomys} rodents on the transmission dynamics of Lassa fever. So, in this work our aim is to investigate the role of predictable variability in time-dependent per capita birth rate of \textit{mastomys} rats/rodents on the transmission dynamics of Lassa fever and to explore factors that contribute to continuous outbreak and how those factors can be curtailed in the presence of one or many intervention strategies that we will evaluate.\\

This paper is planned as follows; Model formulation and description are presented in section 2. In section 3, we displayed the Lassa fever, non-autonomous periodically-forced model. In Sections 4, qualitative analysis of the periodically-forced Lassa fever model is discussed. In section 5, we present the existence of the endemic equilibrium point for the Lassa fever periodically-forced model.  In section 6, we present the evaluation of the Lassa fever intervention strategies using the elasticity of the static quantity, while in section 7, numerical results were shown and conclusions and recommendations of the work are presented in section 8.

\section{Model Formulation and Analysis}
We propose a Lassa fever model with $S_HE_HA_HI_HR_H + S_RI_R$ compartmental design with standard incidence and variable total humans and \textit{mastomys} rat/rodent population.$S_H$ stands for the number of susceptible humans in the population,$E_H$ is referred to as the number of exposed humans in the population, $A_H$ represents the number of asymptomatic humans in the population, $I_H$ is the the number of infected and infectious humans,and the number of recovered humans is $R_H$.The total humans population at time $t$, is given by $N_H(t)$ and it is grouped into Susceptible humans $(S_H(t))$, Exposed humans $(E_H(t))$,Asymptomatic human $(E_H(t))$  Infected humans $(I_H(t))$, and Temporary immune/Recovered humans  $(R_H(t))$. Therefore,
$$N_H(t) = S_H(t) + E_H(t) + A_H(t) + I_H(t) + R_H(t)$$
The total \textit{mastomys natalensis} rat population at time $t$ is given by $N_R(t)$, and it is subdivided into susceptible \textit{mastomys natalensis} rat , $(S_{R}(t))$, and Infected and infectious \textit{mastomys natalensis} rat $(I_{R}(t))$, hence $$N_{R}(t) = S_{R}(t) + I_{R}(t).$$
The population of susceptible humans is produced by the humans per capita birth rate at $\Lambda_h$ followed by the rate at which humans looses their immunity at a rate $\sigma$.  It is reduced by infection, following contacts with infected \textit{mastomys natalensis} rat at a rate $\beta_H$, it is also reduced by infection, following contacts with asymptomatic humans at a rate $\beta_1$ and by infection, following contacts with symptomatic infected humans at a rate $\beta_2$. It is further reduced by the natural death rate of the humans population at a rate $\mu_{H}$.\\
Thus, the rate of change of the population of susceptible humans is given by;
$$\frac{dS_{H}}{dt}  =  \Lambda_H -  \frac{ \beta_{H} S_{H} I_{R}}{N_{H}} - \frac{\beta_{1}A_{H} S_{H}}{N_{H}} - \frac{\beta_{2} I_{H} S_{H}}{N_{H}} + \sigma R_{H}- \mu_{H}S_{H}$$
And the rate of change of the population of exposed humans is given by;
$$\frac{dE_{H}}{dt}  =  \frac{ \beta_{H} S_{H} I_{R}}{N_{H}} + \frac{\beta_{1}A_{H} S_{H}}{N_{H}} + \frac{\beta_{2} I_{H} S_{H}}{N_{H}} - (\theta + \mu_{H}) E_{H} $$
where $\theta$ is rate of progression from exposed to asymptomatic humans  and to infected humans and natural death rate is $\mu_{H}$.\\
The population of asymptomatic humans is generated following the rate at which exposed humans progresses to the asymptomatic compartment at probability of an exposed individual becoming asymptomatic cases upon infection $(1-\upsilon)\theta$, and reduces by the transmission rate from infected rat to asymptomatic humans at a rate $\beta_{3}$ and decreases by recovery at a rate $\gamma_{1}$, it also decreases by Lassa fever induced death at a rate $\omega_1$ and natural death rate $\mu_{H}$. \\
Therefore, the rate of change of the population of asymptomatic humans is given by;
$$\frac{dA_{H}}{dt}  =  (1-\upsilon)\theta E_{H} - \frac{\beta_{3} A_{H} I_{R}}{N_{H}} - (\omega_1 + \gamma_{1} + \mu_{H}) A_{H}$$
The infected humans population is defined as:
$$\frac{dI_{H}}{dt}  =  \upsilon\theta E_{H} + \frac{\beta_{3} A_{H} I_{R}}{N_{H}} -(\omega_2 + \gamma_{2} + \mu_{H}) I_{H}$$
where $\theta\upsilon$ is the probability of an exposed individual becoming symptomatic upon infection, and it decreases by the transmission rate from infected rat to asymptomatic humans at a rate $\beta_{3}$, $\omega_2$ is the Lassa fever induced death rate in the infectious and symptomatic class, while $\gamma_{2}$ is the recovery rate of the the infected and natural death rate is $\mu_{H}$.\\
The population of the recovery/temporary immune is generated by the recovery rate $\gamma_{1}$ of the asymptomatic humans and the the recovery rate of the infected and symptomatic humans. It is reduced by the loss of immunity at a rate $\sigma$ and natural death at a rate $\mu_{H}$\\
Therefore, the rate of change of the recovered population is given by;
$$\frac{dR_{H}}{dt}  = \gamma_{1} A_H + \gamma_{2} I_{H} - (\sigma + \mu_{H}) R_{H}$$
The population of susceptible \textit{mastomys natalensis} rat is generated by the time-dependent per capita birth rate of \textit{mastomys natalensis} rat, $\Lambda_R(t)$ $\Lambda_R(t) =\Lambda_{0}(1+\Lambda_{1}\cos(\frac{2\pi}{365.25}(t+\phi)))$. It is reduced by Infection, following number of contact with an infected \textit{mastomys natalensis} rat at a rate $\beta_{R}$ It is further reduced by the natural death rate of the susceptible \textit{mastomys natalensis} rat population at a rate $\mu_{R}$.\\
Therefore, the rate of change of the population of susceptible \textit{mastomys natalensis} rat/rodent is given by
$$\frac{dS_{R}}{dt}  =  \Lambda_{R}(t)N_{R}(1-\frac{N_R I_R}{M}) - \beta_{R} S_{R} I_{R}- \mu_{R} S_{R}$$
Hence, the rate of change of the population of Infected \textit{mastomys natalensis} rat is given by
$$\frac{dI_{R}}{dt}  =   \beta_{R} S_{R} I_{R}- \mu_{R} I_{R}.$$
where $\mu_R$ is the natural death rate of \textit{mastomys natalensis} rat.

The model schematic diagram is given in figure 1 below:\\
\newpage
\tikzstyle{line} = [draw, -latex']
\tikzstyle{block} = [rectangle, draw, fill=#1,
    text centered, solid, thin,  minimum size=2em, node distance = 4cm]
\tikzstyle{init} = [pin edge={to-,thin,black}]
 \tikzstyle{xs} = [xshift=#1mm]
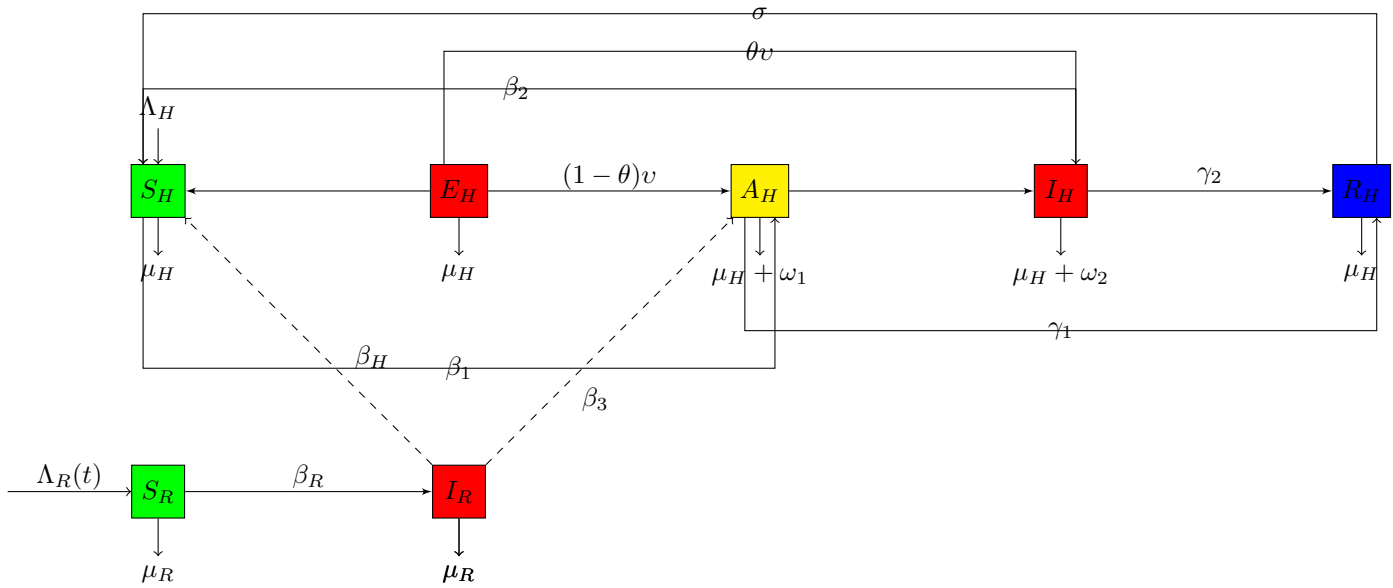
\begin{figure}
\begin{tikzpicture}
\node [block=green, pin={[init]above:$\Lambda_H$}] (sH) {$S_H$};
%\node [block=green] (sh) {$S_h$};
\node [block=green, below of =sH] (sR) {$S_R$};
\node [block=red, right of =sH] (eH) {$E_H$};
\node [block=yellow, right of =eH] (aH) {$A_H$};
\node [block=red, right of =aH] (iH) {$I_H$};
\node [block=blue, right of =iH] (rH) {$R_H$};
\node [block=red, right of =sR] (iR) {$I_R$};
\node (b) [left of =sR,node distance=2cm, coordinate] {$S_R$};
\node [coordinate] (end) [right of=rH, node distance=1cm]{};
%pin={[init]above:$v_0$}]
%arrows
\path[line](eH)--node[yshift=0.2cm] {}(sH);
%\path[<-] (sh) edge node {$vm$} (ih);
\path[line](eH)--node[yshift=0.2cm] {$(1-\theta)\upsilon$}(aH);
\path[line](aH)--node[yshift=0.2cm] {}(iH);
\path[line](iH)--node[yshift=0.2cm] {$\gamma_{2}$}(rH);
\path[draw,->] ([xs=-2] eH.north) -- ++ (0,1.5)
                                -| ([xs=2] iH.north) node[pos=0.25]{$\theta \upsilon$};
\path[draw,->] ([xs=-2] aH.south) -- ++ (0,-1.5)
                                -| ([xs=2] rH.south) node[pos=0.25]{$\gamma_{1}$};
\path[draw,->] ([xs=-2] sH.south) -- ++ (0,-2.0)
                                -| ([xs=2] aH.south) node[pos=0.25]{$\beta_{1}$};
\path[draw,<-] ([xs=-2] sH.north) -- ++ (0,1.0)
                                -| ([xs=2] iH.north) node[pos=0.2]{$\beta_{2}$};
\path[draw,->] (iR) edge [dashed,pos=0.35,"$\beta_H$" '] (sH)
               (iR) edge [dashed,pos=0.35,"$\beta_3$" '] (aH)
                (iR.south) -- ++ (0,-0.5) node[below] {$\mu_R$};
%\path[draw,<-] (sh.south) -- ++ (0,8mm) node[above] {$\Lambda_h$};
\path[line](sR)--node[yshift=0.2cm] {$\beta_R$}(iR);
\path[draw,<-] ([xs=-2] sH.north) -- ++ (0,2.0)
                                -| ([xs=2] rH.north) node[pos=0.25]{$\sigma$};
\path[draw,->](sH.south) -- ++ (0,-0.5) node[below] {$\mu_H$};
\path[draw,->](eH.south) -- ++ (0,-0.5) node[below] {$\mu_H$};
\path[draw,->](aH.south) -- ++ (0,-0.5) node[below] {$\mu_H + \omega_1$};
\path[draw,->](iH.south) -- ++ (0,-0.5) node[below] {$\mu_H+\omega_2$};
\path[draw,->](rH.south) -- ++ (0,-0.5) node[below] {$\mu_H$};
%\path[draw,->](ah.south) -- ++ (0,-0.5) node[below] {$\mu_h$};
\path[draw,->](sR.south) -- ++ (0,-0.5) node[below] {$\mu_R$};
\path[draw,->](iR.south) -- ++ (0,-0.5) node[below] {$\mu_R$};
\path[->] (b) edge node[yshift=0.2cm] {$\Lambda_{R}(t)$} (sR);
%\path[draw,->] (a0 |- sv) to["$\Lambda_{v}(t)$"] (sv);
%\path[draw,->] (ah) edge node {$\mu_h$} ;
%\path[draw, <-] (sh) -- +(0,1.5) -| node[near start] {$k+vm$} (ah);
%\path[line,dashed](ah) --node[yshift=-0.2cm, xshift=-0.4cm]{$\beta_{h1}$} (iv);
\end{tikzpicture}
\caption{Flowchart for the Lassa fever model.}
\end{figure}

The associated model variables and parameters are described in Tables 1. and 2. respectively.
\begin{table}[H]
\caption{\small Description of the state variables in the Lassa-fever model} \centering \small
\begin{tabular}{cc}\hline
Variable&Description \\ \hline
$S_H(t)$ &Number of Susceptible, naive and non-immune humans host at time ,$t$ \\
$E_H(t)$ &Number of Exposed humans host at time, $t$ \\
$A_H(t)$ &Number of Infectious,Asymptomatic humans host at time, $t$ \\
$I_H(t)$ &Number of Infected, infectious, clinically ill, and symptomatic humans host at time, $t$ \\
$R_H(t)$ &Number of Recovered/Temporary immune humans host at time, $t$ \\
$S_R(t)$ &Number of Susceptible \textit{Mastomys natalensis} Rat(Rodent) at time, $t$ \\
$I_R(t)$ &Number of Infected \textit{Mastomys natalensis} Rat(Rodent) at time, $t$\\\hline
\end{tabular}
\label{tabmx1}
\end{table}

\begin{table}[H]
\caption{\small Table showing parameters of the Lassa fever model}
\centering
\small
\begin{tabular}{lccccl}\hline
Parameter&Description \\ \hline
$\Lambda_{H}$&Recruitment rate of humans. $humans \times Time^{-1}$\\
$\Lambda_{R}(t)$&time-dependent per capita birth rate of \textit{mastomys natalensis} rat. $Time^{-1}$\\
$\Lambda_{0}$&the baseline per capita birth rate of \textit{mastomys natalensis} rat. $Time^{-1}$\\
$\Lambda_{1}$& amplitude or degree of seasonality. $Time^{-1}$\\
$\phi$& phase angle (month of peak in seasonal forcing). $Time^{-1}$\\
$\beta_H$&Infection rate from \textit{mastomys natalensis} rat-to-susceptible humans. $Time^{-1}$\\
$\beta_1$&Infection rate from infected humans-to-susceptible humans. $Time^{-1}$\\
$\beta_2$&Infection rate from asymptomatic humans-to-susceptible humans. $Time^{-1}$\\
$\beta_{3}$&Transmission rate from infected rat to asymptomatic humans.$Time^{-1}$\\
$\beta_R$&Infection rate from infected \textit{mastomys natalensis} rat-to-susceptible \textit{mastomys natalensis} rat. $Time^{-1}$\\
$\sigma$&rate at which temporary immune humans looses their immunity. $Time^{-1}$\\
$\gamma_{1}$&recovery rate of infected humans. $Time^{-1}$\\
$\gamma_{2}$&recovery rate of asymptomatic humans. $Time^{-1}$\\
$\omega_1$&Lassa fever-induced death rate of symptomatic. $Time^{-1}$\\
$\omega_2$&Lassa fever-induced death rate of asymptomatic infected humans. $Time^{-1}$\\
$\theta\upsilon$&Probability of an exposed individual becoming symptomatic upon infection. $Time^{-1}$\\
$(1-\theta)\upsilon$&Probability of an exposed individual becoming asymptomatic cases upon infection. $Time^{-1}$\\
$M$&carrying capacity of the \textit{mastomys natalensis} rat population. $Time^{-1}$\\
$\mu_H$&Natural death rate of humans.$Time^{-1}$\\
$\mu_R$&Natural death rate of \textit{mastomys natalensis} rat. $Time^{-1}$\\\hline
\end{tabular}
\label{tabmx1}
\end{table}

\section{The Lassa-fever Non-Autonomous periodically-forced Model}
The following is obtained from the model descriptions;
\begin{equation}
\begin{array}{lcl}
\ \\
\frac{dS_{H}}{dt} & = & \Lambda_H -  \frac{ \beta_{H} S_{H} I_{R}}{N_{H}} - \frac{\beta_{1}A_{H} S_{H}}{N_{H}} - \frac{\beta_{2} I_{H} S_{H}}{N_{H}} + \sigma R_{H}- \mu_{H}S_{H}\\
\ \\
\frac{dE_{H}}{dt} & = & \frac{\beta_{H} S_{H} I_{R}}{N_{H}} + \frac{\beta_{1}A_{H} S_{H}}{N_{H}} + \frac{\beta_{2} I_{H} S_{H}}{N_{H}} - (\theta + \mu_{H}) E_{H} \\
\ \\
\frac{dA_{H}}{dt} & = & (1-\upsilon)\theta E_{H} -\frac{\beta_{3} A_{H} I_{R}}{N_{H}} - (\omega_{1} + \gamma_{1} + \mu_{H}) A_{H} \\
\ \\
\frac{dI_{H}}{dt} & = & \upsilon \theta E_{H} + \frac{\beta_{3} A_{H} I_{R}}{N_{H}} - (\omega_{2} + \gamma_{2} + \mu_{H}) I_{H} \\
\ \\
\frac{dR_{H}}{dt} & = & \gamma_{1} A_H + \gamma_{2} I_{H} - (\sigma + \mu_{H}) R_{H} \\
\ \\
\frac{dS_{R}}{dt} & = & \Lambda_{R}(t)N_{R}(1-\frac{N_R}{M}) - \beta_{R} S_{R} I_{R}- \mu_{R} S_{R}\\
\ \\
\frac{dI_{R}}{dt} & = & \beta_{R} S_{R} I_{R} - \mu_{R} I_{R}\\
\ \\
\frac{dN_{H}}{dt} & = &\Lambda_H  - \mu_{H} N_{H} - \omega_{1} A_H - \omega_{2} I_H) \\
\ \\
\frac{dN_{R}}{dt} & = & \Lambda_{R}(t)N_{R}(1-\frac{N_R}{M}) - \mu_{R} N_{R}.
\ \\
\end{array}
\end{equation}
%Lambda0*(1+Lambda1*cos(2*pi*(t-phi)/12))
where \\
$$\Lambda_{R}(t) = \Lambda_{0}(1+\Lambda_{1}\cos(\frac{2\pi}{365.25}(t+\phi)))$$
subject to the initial conditions
$$S_{H}(0) = S_{H,0}, \; E_{H}(0) = E_{H,0}, \; A_{H}(0) = A_{H,0}, \; I_{H}(0) = I_{H,0}, \; R_H(0) = R_{H,0}, \; S_{R}(0) = S_{R,0}, \; I_{R}(0) = I_{R,0}.$$
We discuss in details the basic properties of the periodically-forced Lassa-fever model $(1)$ which entails its basic mathematical analysis in the next section.\\
Suppose $\Lambda_{R}(t)$ denotes the average, given that:
$$\Lambda_{R}(t)= \frac{1}{Q} \int_{t_{0}}^{t+Q} \Lambda_{R}(s) ds$$, $t \geq 0$ and we assume $\Lambda_{R}(t)>0$
The basic properties of the Lassa-fever model $(1)$ which entails its basic mathematical analysis is presented in the next section.
\section{Qualitative Analysis of the periodically-forced Lassa fever model}
\subsection{Positivity of Solutions}
We found it pertinent to prove that all the state variables of the periodically-forced Lassa fever transmission dynamic model $(1)$ are non-negative at all time $(t)$ for the model to be epidemiologically meaningful and mathematically well-posed.\\
\textbf{Theorem 1.}: Given that the initial data $S_{H} \geq 0$, $E_{H}\geq 0$, $A_{H} \geq 0$, $I_{H}\geq 0$, $R_{H}\geq0$,$S_{R}\geq 0$, $I_{R} \geq 0$ then the solutions $(S_{H},E_{H},A_{H}, I_{H}, R_{H},S_{R},I_{R})$ of the periodically-forced Lassa fever model $(1)$ are non-negative for all $t>0$.\\
Hence,\\
$\limsup_{t\rightarrow \infty} N_{H}(t) \leq \frac{\Lambda_{H}}{\mu_{H}}$, $\&$ $\limsup_{t\rightarrow \infty} N_{R}(t) \leq \frac{ M(\Lambda_R(t) - \mu_R)}{\Lambda_R(t)}$.\\
Such that\\
$N_{H}=S_{H}+E_{H}+A_{H}+I_{H}$ $\&$ $N_{R}=S_{R}+I_{R}$.\\
\textbf{Proof.}: Suppose $\Omega_{L}= \sup{t>0 :S_{H}(t)>0,E_{H}(t)>0,A_{H}(t)>0, I_{H}(t)>0, R_{H}(t)>0,S_{R}(t)>0,I_{R}(t)>0}$. Since $S_{H}(t)>0$,$E_{H}(t)>0$,$A_{H}(t)>0$, $I_{H}(t)>0$, $R_{H}(t)>0$,$S_{R}(t)>0$,$I_{R}(t)>0$, then $\Omega_{L}>0$. If $\Omega_{L}>\infty$, then $S_{H},E_{H},A_{H}, I_{H}, R_{H},S_{R},I_{R}$ equals zero at $\Omega_{L}$.\\
It follows from the first equation of the system (1) that\\
$\frac{dS_{H}}{dt} =  \Lambda_H -  \frac{ \beta_{H} S_{H} I_{R}}{N_{H}} - \frac{\beta_{1}A_{H} S_{H}}{N_{H}} - \frac{\beta_{2} I_{H} S_{H}}{N_{H}} + \sigma R_{H}- \mu_{H}S_{H}$\\
Hence,\\
$\frac{d}{dt}{S_{H}(t)exp[(\beta_{H}I_{R}+\beta_{1}A_{H}+\beta_{2} I_{H}+\mu_{H})t]}=\int_0^{\Omega_{L}}(\Lambda_{H} + \sigma R_{H})exp[(\beta_{H}I_{R}+\beta_{1}A_{H}+\beta_{2} I_{H}+\mu_{H})q]dq$
Such that\\
$S_{H}(\Omega_{L}) = S_{H}(0)exp[-(\beta_{H}I_{R}+\beta_{1}A_{H}+\beta_{2} I_{H}+\mu_{H})\Omega_{L}]+exp[-(\beta_{H}I_{R}+\beta_{1}A_{H}+\beta_{2} I_{H}+\mu_{H})\Omega_{L}] \times \int_0^{\Omega_{L}}(\Lambda_{H} + \sigma R_{H})exp[(\beta_{H}I_{R}+\beta_{1}A_{H}+\beta_{2} I_{H}+\mu_{H})q]dq$

and\\

$A_{H}(\Omega_{L}) = A_{H}(0)exp[-(\frac{\beta_{H}I_{R}}{N_{H}}+(\omega + \gamma_{1} + \mu_{H})\Omega_{L}+exp[(\frac{\beta_{H}I_{R}}{N_{H}}+\omega + \gamma_{1} + \mu_{H})\Omega_{L}])q]dq>0$\\
It can as well be proved that $E_{H}>0$, $I_{H}>0$, $R_{H}>0$, $S_{R}>0$, $I_{R}>0$ $\forall$ $t>0$.\\
\textbf{Remark 1}: The solutions $(S_{H},E_{H},A_{H}, I_{H}, R_{H},S_{R},I_{R})$ of the Lassa fever autonomous model with non-negative initial data will remain non-negative every time as $t>0$.

\subsection{Boundedness of the solution}
\textbf{Theorem 2.}: Every solution $(S_{H},E_{H},A_{H}, I_{H}, R_{H},S_{R},I_{R})$ of the Lassa fever non-autonomous model are bounded. Thus, if $\limsup_{t\rightarrow \infty} N_{H}(t) \leq \frac{\Lambda_{H}}{\mu_{H}}$, $\&$ $\limsup_{t\rightarrow \infty} N_{R}(t) \leq M$.
Such that\\ $N_{H}=S_{H}+E_{H}+A_{H}+I_{H}$ $\&$ $N_{R}=S_{R}+I_{R}$.\\
\textbf{Proof.}: Suppose $0 < I_{H}(t) \leq N_{H}(t)$, $0<A_{H}(t) \leq N_{H}(t)$  and $0<I_{R}\leq N_{R}(t)$.Hence, the sum of the first five equations of the model (1) is given by
$$\frac{dN_{H}(t)}{dt} = \Lambda_{H} - \mu N_{H} - \omega_{1} A_{H} - \omega_{2} I_{H}$$
while the sum of the last two equations of the model $(1)$ is given by
$$\frac{dN_{R}}{dt} = \Lambda_{R}(t) N_{R}(1-\frac{N_R}{M}) - \mu_{R} N_{R}.$$
Therefore,\\
$\Lambda_{H} - (\mu_{H} + \omega_{1} + \omega_{2})N_{H}(t) \leq \frac{dN_{H}(t)}{dt}\leq \Lambda_{H} - \mu_{H}$\\
$\Lambda_{R}(t)N_{R}(1-\frac{N_R}{M}) - \mu_{R} N_{R}\leq \frac{dN_{H}(t)}{dt}\leq\Lambda_{R}(t) N_{R}(1-\frac{N_R}{M}) - \mu_{R} N_{R}$\\
Therefore,\\
$\frac{\Lambda_{H}}{\mu_{H}+\omega_{1}+\omega_{2}}\leq \liminf_{t\rightarrow \infty} N_{H}(t) \leq \limsup_{t\rightarrow \infty} N_{H}(t) \leq \frac{\Lambda_{H}}{\mu_{H}}$\\
and \\
$\frac{ M(\Lambda_R(t) - \mu_R)}{\Lambda_R(t)} \leq \liminf_{t\rightarrow \infty} N_{R}(t) \leq \limsup_{t\rightarrow \infty} N_{R}(t) \leq \frac{ M(\Lambda_R(t) - \mu_R)}{\Lambda_R(t)}$.\\
\textbf{Remark 2}: This shows that the periodically-forced Lassa fever model $(1)$ is epidemiologically meaningful and mathematically well posed in the region $J = J_{H} \cup J_{R} \subset \mathbb{R}^{5}_{+} \times \mathbb{R}^{2}_{+}$.Hence, the total population of humans and the \textit{mastomys natalensis} rats is bounded above and below.\\
\textbf{Theorem 3.}:The region $J = J_{H} \cup J_{R} \subset \mathbb{R}^{5}_{+} \times \mathbb{R}^{2}_{+}$ is positively-invariant for the model $(1)$ with non-negative initial condition in $\mathbb{R}^{7}_{+}$.\\
\textbf{Proof.}: The Lassa fever non-autonomous model $(1)$ is analysed in a biologically feasible region as follows: the model equation $(1)$ is divided into human population compartment $N_{H}$ and the \textit{Mastomys natalensis} rat population $N_{R}$. Hence, we consider the feasible region:\\
$J = J_{H} \cup J_{R} \subset \mathbb{R}^{5}_{+} \times \mathbb{R}^{2}_{+}$;
with \\
$J_{H}=\{ (S_{H},E_{H},A_{H}, I_{H}, R_{H})\in \mathbb{R}^{5}_{+} : S_{H}+E_{H}+A_{H}+I_{H}+R_{H} \leq \frac{\Lambda_{H}}{\mu_{H}}\}$, \\
and \\
$J_{R} = \{ (S_{R},I_{R}) \in \mathbb{R}^{2}_{+} : S_{R}+I_{R}\leq \frac{ M(\Lambda_{R}(t)- \mu_R)}{\Lambda_{R}(t)}\}$.\\
We established the positive invariance of $J$ which means solutions in $J$ remains in $J$ for every $t>0$. The rate of change of the humans and the \textit{Mastomys natalensis} rats population is given by\\
$\frac{dN_{H}(t)}{dt} \leq \Lambda_{H} - \mu N_{H}$\\
$\&$ $\frac{dN_{R}}{dt} \leq \frac{ M(\Lambda_{R}(t)- \mu_R)}{\Lambda_{R}(t)}$.\\
We then applied the standard comparison theorem by Lakshmikantham et al. (1989) to prove that\\
$N_{H}(t) \leq N_{H}(0)e^{-\mu_{h}t} + \frac{\Lambda_{H}}{\mu_{H}}(1-e^{(-\mu_{h}t)}$ and $N_{R}(t) \leq \frac{ M(\Lambda_R(t) - \mu_R)}{\Lambda_R(t)}\} $.
In particular, \\
$N_{H}(t)\leq \frac{\Lambda_{H}}{\mu_{H}}$ $\&$ $N_{R}(t) \leq \frac{ M(\Lambda_R(t) - \mu_R)}{\Lambda_R(t)}\}$\\
whenever\\
$N_{H}(0)\leq \frac{\Lambda_{H}}{\mu_{H}}$ $\&$ $N_{R}(0) \leq \frac{ M(\Lambda_R(t) - \mu_R)}{\Lambda_R(t)}\}$.\\
Hence, $J$ is the region that is positively-invariant and it is sufficient to consider the dynamics of the flow generated by $(1)$ in $J$.\\
\textbf{Remark 3}: $J$ is the region where the model $(1)$ is epidemiologically meaningful and mathematically well-posed. Therefore, all solutions of the model $(1)$ with initial condition in $J$ stay in $J$ at all time $t>0$.\\
\subsection{Equilibrium solution and Analysis of the Lassa fever non-autonomous model}
The model equation $(1)$ is analysed in this section so that we can obtain the equilibrium points of the system.
In order to provide an answer for the long term behaviour of the Lassa fever autonomous model, we obtain the equilibrium solution of the model $(1)$ by setting the equations of model $(1)$ to zero, with\\ $\frac{dS_{H}}{dt}=\frac{dE_{H}}{dt}=\frac{dA_{H}}{dt}=\frac{dI_{H}}{dt}=\frac{dR_{H}}{dt}=\frac{dS_{R}}{dt}=\frac{dI_{R}}{dt}=0$.\\
We obtain the values of the variables denoted as $S^{\ast}_{H}$,$E^{\ast}_{H}$,$A^{\ast}_{H}$,$I^{\ast}_{H}$,$R^{\ast}_{H}$,$S^{\ast}_{R}$,$I^{\ast}_{R}$ which satisfy this criteria. It shows that there will be no trivial equilibrium and that the population will never goes into extinction if the birth rate of humans $\Lambda_{H}$ and the birth rate of \textit{mastomys natalensis} rat  $\Lambda_{R}$ are non-zero. Therefore, $E_{H}=A_{H}=I_{H}=R_{H}=0$ when the Lassa fever disease is not present in the population, the model equation $(1)$ has a steady state, $J_{0}$ which is the disease-free equilibrium (DFE). Hence, the DFE of the model is given by
$$J_{0}=(S^{\ast}_{H}=\frac{\Lambda}{\mu},E^{\ast}_{H}=0,A^{\ast}_{H}=0,I^{\ast}_{H}=0,R^{\ast}_{H}=0,S^{\ast}_{R}=\frac{\Lambda_{R}(t) N_{R}(t)(M-N_{R})}{M \mu_{R}},I^{\ast}_{R}=0)$$

\subsection{The Basic Reproduction number $R_{L}$ for the Lassa fever autonomous model}
The reproduction number is a very important threshold quantity in epidemiology which measures the average number of new cases in a completely susceptible population. We used the next generation matrix method to calculate the $R_{L}$. The next generation method is the spectral radius of the next generation matrix (Van den Dressche $\&$ Watmough (2002)).\\
\textbf{Proposition 1.}: Suppose the $R_{L}$ of the Lassa fever autonomous model $(1)$ which is computed as the largest positive eigenvalue of the next generation matrix is given by\\ $R_L=\frac{\beta_R \theta \mu_H (\beta_{1}(1-\upsilon)+\upsilon \beta_2)-\beta_R \beta_1 \theta (1-\upsilon)(\omega_2 + \gamma_2)+\upsilon\beta_2\beta_R \theta(\gamma_1 + \omega_1)}{\mu^{2}_{H} \mu_{R}(\gamma_1 + \omega_1 + \mu_{H})(\theta+\mu_{H})(\omega_2+\gamma_2+\mu_H)}$.\\
\textbf{Proof:}
generation matrix. Suppose
\begin{equation}
  \mathfrak{F}  = \left(
   \begin{array}{cccc}
         0 &  \frac{\beta_{1} \Lambda_{H}}{N_{H} \mu_{H}} &  \frac{\beta_{2} \Lambda_{H}}{N_{H} \mu_{H}} &  \frac{\beta_{1} \Lambda_{H}}{N_{H} \mu_{H}} \\
         0 &  0 &  0 & 0 \\
         0 &  0 &  0 & 0\\
         0 &  0&  0& \frac{\beta_{R} \Lambda_{R}(t)(M-N_{R})}{M \mu_{R}}
       \end{array}
       \right)
\end{equation}
The rate of appearance of the disease in a compartment is $\mathfrak{F}$ while $\mathfrak{V}$ is the transfer of individuals and \textit{mastomys natalensis} rat into one compartment. Hence $\mathfrak{F}$ is a Jacobian matrix evaluated at $J_{0}$ and the Jacobian matrix of $V$ is evaluated at $J_{0}$ which yields
\begin{equation}
   V  = \left(
   \begin{array}{ccccc}
         (\theta + \mu_{H}) &  0 &  0 & 0 \\
         -(1-\upsilon)\theta &   (\omega_{1} + \gamma_{1} + \mu_{H})  &  0 &  0 \\
         -\upsilon \theta &  0 & (\omega_{2} + \gamma_{2} + \mu_{H})  &  0\\
         0 &  0 &  0 & \mu_{R}
       \end{array}
       \right)
\end{equation}

\begin{equation}
   V^{-1}  = \left(
   \begin{array}{cccc}
         \frac{1}{(\theta + \mu_{H})}  &  0 &  0 & 0 \\
         \frac{(1-\upsilon)\theta }{(\theta + \mu_{H})(\omega_{1} + \gamma_{1} + \mu_{H})} & \frac{1}{(\omega_{1} + \gamma_{1} + \mu_{H})} & 0  &  0 \\
         \frac{\upsilon \theta }{(\theta + \mu_{H})(\omega_{2} + \gamma_{2} + \mu_{H})} & 0 & \frac{1}{(\omega_{2} + \gamma_{2} + \mu_{H})} &  0\\
         0 &  0 &  0 & \frac{1}{\mu_{R}}
       \end{array}
       \right)
\end{equation}

The product of matrix $\mathfrak{F}$ and
$V^{-1}$ gives
\begin{equation}
  {\mathfrak{F} V^{-1}} = \left(
         \begin{array}{ccccc}
        V_{1} & \frac{\beta_{1} \Lambda_{H}}{\mu_{H} N_{H}(\omega_{1} + \gamma_{1} + \mu_{H})} & \frac{\beta_{2} \Lambda_{H}}{\mu_{H} N_{H}(\omega_{2} + \gamma_{2} + \mu_{H})} &  \frac{\beta_{H} \Lambda_{H}}{\mu_{H} N_{H} \mu_{R}} & \\
         0 &  0  & 0 &  0 \\
         0 &  0 & 0  &  0\\
         0 &  0 &  0 & \frac{\beta_{R} \Lambda_{R}(t) (M-N_{R})}{M \mu^{2}_{R}}
       \end{array}
       \right)
\end{equation}

Let\\
$V_{1}=  \frac{\beta_{1} \Lambda_{H} (1-\upsilon)\theta}{\mu_{H} N_{H}(\theta+\mu_{H})(\omega_{1} + \gamma_{1} + \mu_{H})}+\frac{\beta_{2} \Lambda_{H} \upsilon \theta}{\mu_{H} N_{H}(\theta+\mu_{H})(\omega_{2} + \gamma_{2} + \mu_{H})}$\\
Therefore the eigenvalue of $\mathfrak{F} V^{-1}$ is calculated to obtain $R_{L}$ for the Lassa fever autonomous model $(1)$ is given by
$$\langle \Lambda_{R}\rangle= \frac{1}{Q} \int_{t_{0}}^{t+Q} \Lambda_{R}(s) ds$$,$t \geq 0$
\begin{equation}
R_{L_{1}} = \frac{ \theta \mu_H (\beta_{1}(1-\upsilon)+\upsilon \beta_2)-\beta_1 \theta (1-\upsilon)(\omega_2 + \gamma_2)+\upsilon\beta_2 \theta(\gamma_1 + \omega_1)}{N_{H} \mu_{H} (\gamma_1 + \omega_1 + \mu_{H})(\theta+\mu_{H})(\omega_2+\gamma_2+\mu_H)}
\end{equation}
\begin{equation}
R_{L_{2}}=\frac{(\langle \Lambda_{R}\rangle-\mu_{R})\beta_{R} M}{\langle \Lambda_{R}\rangle \mu_{R} N_{R}}
\end{equation}
The Basic reproduction number according to Van den Driessche $\&$ Watmough (2008) is given by;
\begin{equation}
R_{L} = \sqrt{\frac{( \Lambda_{R}-\mu_{R})\beta_{R} M(\theta \mu_H (\beta_{1}(1-\upsilon)+\upsilon \beta_2)-\beta_1 \theta (1-\upsilon)(\omega_2 + \gamma_2)+\upsilon\beta_2 \theta(\gamma_1 + \omega_1))}{\Lambda_{R}\mu_{R} N_{R} N_{H} \mu_{H} (\gamma_1 + \omega_1 + \mu_{H})(\theta+\mu_{H})(\omega_2+\gamma_2+\mu_H)}}
\end{equation}
While the average basic reproduction number in the presence of the periodic function is given by;
\begin{equation}
\tilde{R}_{L} = \sqrt{\frac{ (\langle \Lambda_{R}\rangle-\mu_{R})\beta_{R} M(\theta \mu_H (\beta_{1}(1-\upsilon)+\upsilon \beta_2)-\beta_1 \theta (1-\upsilon)(\omega_2 + \gamma_2)+\upsilon\beta_2 \theta(\gamma_1 + \omega_1))}{\langle \Lambda_{R}\rangle \mu_{R} N_{R} N_{H} \mu_{H} (\gamma_1 + \omega_1 + \mu_{H})(\theta+\mu_{H})(\omega_2+\gamma_2+\mu_H)}}
\end{equation}
It is essential to know that $\tilde{R}_{L}\neq R_{L}$\\
\textbf{Remark 4}: This shows that persistence and/or extinction of Lassa fever is obtained by calculating the $\tilde{R}_{L}$ of the Lassa fever such that if $\tilde{R}_{L} < 1$, the disease will extinct from the population and if $\tilde{R}_{L} > 1$, the disease will persist in the population. We discuss in the section the existence of the disease-free equilibrium.
\subsection{Existence of the Disease Free Equilibrium (DFE) of the Lassa fever non-autonomous model}
In a situation without the Lass fever, the Lassa fever free equilibrium of the Lassa fever autonomous model $(1)$ becomes:\\
$J_{0}=(S^{\ast}_{H},E^{\ast}_{H},A^{\ast}_{H},I^{\ast}_{H},R^{\ast}_{H},S^{\ast}_{R},I^{\ast}_{R})$ such that  $E^{\ast}_{H}=A^{\ast}_{H}=I^{\ast}_{H}=R^{\ast}_{H}=I^{\ast}_{R}=0$,
This yields,
$$J_{0}=(S^{\ast}_{H},0,0,0,0,S^{\ast}_{R},0)$$
When we subject each of the equation in the model $(1)$ to zero it yields;\\
$0=\Lambda_H -  \frac{ \beta_{H} S_{H} I_{R}}{N_{H}} - \frac{\beta_{1}A_{H} S_{H}}{N_{H}} - \frac{\beta_{2} I_{H} S_{H}}{N_{H}} + \sigma R_{H}- \mu_{H}S_{H}$
therefore,
$S^{\ast}_{H}=\frac{\Lambda_{H}}{\mu_{H}}$

and

$\Lambda_{R}(t)N_{R}(1-\frac{N_R}{M}) - \frac{\beta_{R} S_{R} I_{R}}{N_{R}}- \mu_{R} S_{R}=0$,\\
$\Lambda_{R}(t)N_{R}(1-\frac{N_R}{M}) - \mu_{R} N_{R}=0$,

we obtain,
$S^{\ast}_{R}=\frac{M(\Lambda_{R}(t)-\mu_{R})}{\Lambda_{R}(t)}$

Hence, the Lassa fever free equilibrium is given by\\
$$J_{0}=(\frac{\Lambda_{H}}{\mu_{H}},0,0,0,0,\frac{M(\Lambda_{R}(t)-\mu_{R})}{\Lambda_{R}(t)},0)$$
The \textit{Mastomys Natalensis} rodent compartment is separated from the humans compartment, hence the equation is linear with non-constant coefficients. It can be solved clearly as given below:\\
$G_{R}=G_{R}(0) \exp^{\int^{t}_{0}(\beta_{R}\frac{M}{N_{R}}\frac{\Lambda_{R}(s)-\mu_{R}}{\Lambda_{R}(s)}-\mu_{R}) ds}$\\
We introduce the average of a periodic function over its period in order to define the Lassa fever reproduction number. Obtaining the reproduction number we recall by definition, if $j(t)$ is a periodic function with period $Q$, then the average of $j$ is given by\\
$\langle j \rangle = \frac{1}{Q}\int^{Q}_{0} j(s) ds$\\
By perturbation analysis:\\
$G_{R}(t)=G_{R}(0) \exp^{(\frac{1}{q}\int^{q}_{0}(\beta_{R} M \frac{\Lambda_{R}(q)-\mu_{R}}{\Lambda_{R}(q)}-\mu_{R}) dq)q}$
$\simeq G_{R}(0) \exp^{(\int^{q}_{0}(\beta_{R}M\frac{\langle\Lambda_{R}\rangle-\mu_{R}}{\langle \Lambda_{R} \rangle}-\mu_{R})dq)q}$
for large sufficient $q$, the expression on the RHS goes to $\pm \infty$ if and only if\\ $\beta_{R}\frac{M}{N_{R}}\frac{\langle\Lambda_{R}\rangle-\mu_{R}}{\langle \Lambda_{R} \rangle}-\mu_{R}>0$\\
which prompts us to define the following reproduction number\\
$R_{L_{2}}=\beta_{R}M\frac{\langle\Lambda_{R}\rangle-\mu_{R}}{\langle \Lambda_{R} \rangle}-\mu_{R}$\\
We note that $\Lambda_{R}(t)$ is periodic with period 365.25 days, hence $Q=365.25$ and we summarised the whole concept here in the following result.\\
\textbf{Lemma 1.} : The disease free equilibrium (DFE) is locally asymptotically stable whenever $R_{L_{2}}<1$, then\\
 $G_{R}(t)\rightarrow 0$ and if $R_{L_{2}}>1$, then $|G_{R}(t)\rightarrow \infty$, and the DFE is unstable.\\
\textbf{Lemma 2.}: Suppose the Lassa fever disease (LFD) persists in the community such that \\
$\lim_{p \rightarrow \infty}\frac{1}{p} \int_{0}^{p} I_{R}(t)dt >0$. If $\Lambda_{R} (t) > \mu_{R}$\\
and $\tilde{R}_{L}>1$, and the disease will extinct (which implies that $I_{R}\rightarrow 0$ as $t \rightarrow \infty$, when $\tilde{R}_{L}<1$). Therefore, this result holds for any periodically forced birth rate $\Lambda_{R}(t)$ which satisfies the assumptions that $\Lambda_{R}(t)$ is continuous for all $t\geq 0$ and that $\exists$ a constants $\Lambda^{\min}_{R}$, $\Lambda^{\max}_{R}$ such that $0\leq \Lambda^{\min}_{R} \leq \Lambda^{\max}_{R}$ $\forall$ $t \geq 0$ and $\forall$ $t \geq 0$ the average $\frac{1}{Q} \int_{t_{0}}^{t+Q} \Lambda_{R} (s) ds \rightarrow \tilde{\Lambda_{R}}(t)>0$ as $Q\rightarrow \infty$ uniformly in $t$ i.e. $\forall$ $Q>Q_{0}$ and all $t \geq 0$, $\Lambda_{R}(t)> \mu_{R}$ and finally all the parameters in the model are non-negative.(Ireland et al. (2006))\\
\textbf{Proof:} We prove this theorem by asserting firstly, that for each initial condition in the positive orthant of the population of \textit{Mastomys natalensis} rodent is bounded above and below in consequence we show that in our model, the disease cannot drive the population into extinction. For each initial condition $S_{R}(0)$, $I_{R}(0)$ $\exists$ some positive constants $N^{\min}_{R}$ and $N^{\max}_{R}$ such that $\forall$ $t\geq 0$, $N^{\min}_{R}\leq N_{R}(t) \leq N^{\max}_{R}$. To prove this, we initiate the following notation which says for a positive quantity $Z(t)$, $t \geq 0$, suppose\\
$\overline{Z} = \overline{\lim}_{Q \rightarrow \infty}\int_{0}^{Q}Z(t) dt$, $\underline{Z}=\underline{\lim}_{Q \rightarrow \infty}\int_{0}^{Q}Z(t) dt$, and $\widehat{Z}=\lim_{Q\rightarrow \infty}\int_{0}^{Q}Z(t)dt$\\
Also,\\
$\overline{\phi(Z)} = \overline{\lim}_{Q\rightarrow \infty}\frac{1}{Q}\log\frac{Z(Q)}{Z(0)}$, $\underline{\phi(Z)}=\underline{\lim}_{Q\rightarrow \infty}\frac{1}{Q}\log\frac{Z(Q)}{Z(0)}$, and $\phi(Z)=\lim_{Q\rightarrow \infty}\frac{1}{Q}\log\frac{Z(Q)}{Z(0)}$.\\
Since the total population $N_{R}$ satisfies the differential equation:
$$\frac{dN_{R}}{dt}=(\Lambda_{R}(t)-\mu_{R})N_{R} - \frac{\Lambda_{R}(t) N^{2}_{R}}{M}$$
Let\\
$p(t)=\Lambda_{R}(t)- \mu_{R}$
$$\frac{dN_{R}}{dt}=p(t) N_{R} - \frac{\Lambda_{R}(t) N^{2}_{R}}{M}=(p(t) - \frac{\Lambda_{R}(t) N_{R}}{M})N_{R}$$
Suppose $P_{\max}= \Lambda^{\max}_{R}-\mu_{R}$ we have \\
$\frac{dN_{R}}{dt}\leq 0$ for $N_{R}(t)\geq \frac{P_{\max} M}{\Lambda^{\max}_{R}}$,\\
Let\\
$W=\frac{\Lambda_{R}(t)}{M}$ which implies that $N_{R}(t)\geq \frac{P_{\max}}{W}$.\\
Then $N_{R}(t) \leq \max(N_{R}(0), \frac{P_{\max}}{W})$ $\forall$ $t\geq 0$. Hence, we say $N^{\max}_{R} = \max(N_{R}(0), \frac{P_{\max}}{W})$ such that $N_{R}$
is bounded above. To prove that $N_{R}(t)$ is bounded below, one consider the prevalence of Lassa fever in \textit{Mastomys Natalensis} Rat $f=\frac{I_{R}}{N_{R}}$. \\
The prevalence of Lassa fever in the Rat, $f$ satisfies the differential equation:\\
$$\frac{df}{dt}=(\beta_{R}S_{R}-\mu_{R})f$$.
It is of note that $0\leq f \leq 1$ and $S_{R} \leq N_{R}$, so that for $N_{R} \leq \delta$ we obtain\\
$$\frac{df}{dt}=(\beta_{R}\delta-\mu_{R})f$$
for $\delta > 0$ chosen sufficiently small then the following conditions hold\\
(a) $N_{R}\leq \delta$ (b) $ \beta_{R} \delta - \mu_{R}=- \tau <0$ (c) $\beta_{R} \delta - \mu_{R} > 0$.\\
It is possible to choose $\delta > 0$ such that $N_{R} > 0$, $\mu_{R} > 0$ and $\beta_{R} > 0$.\\
\textbf{Case I} : If we say that $N_{R}(t)$ is not bounded below then $\forall$ $0<m<\delta$ $\exists$ $0 \leq t_{1} < t_{2}$ such that $N_{R}(t_{1}=\delta$ for $t_{1} \leq t \leq t_{2}$.\\
\textbf{Case II}: We show that $N_{R}(t)$ is bounded below. Given $m>0$ for time $t_{2}-t_{1}$ is large sufficiently for the average $\check{p}$ to take over the dynamics of the equation. Such that for any $t_{1} \leq t \leq t_{2}$, $\frac{df}{dt}\leq -\tau f$ holds, so that $f(t) \leq f(t_{1}) e{-\tau (t-t_{1})}$ for $t_{1} \leq t \leq t_{2}$ and the prevalence of Lassa fever decays exponentially.\\
Now, $\frac{dN_{R}(t)}{dt}=(P(t) - W N_{R}) N_{R} \geq (P_{\min} - W \delta) N_{R}$ for $t_{1} \leq t \leq t_{2}$. Suppose $P_{\min} - W \delta \geq 0$ we have $N_{R}(t) \geq N_{R}(t_{1})= \delta$ for $t_{1} \leq t \leq t_{2}$, such that $N_{R}(t_{2}) \geq \delta$ which is a contradiction. Hence, having $P_{\min} - W \delta= - \alpha < 0$. We have that, $N_{R}(t) \geq N_{R}(t_{1}) e^{-\alpha(t-t_{1})}$ for $t_{1} \leq t \leq t_{2}$ such that $m=N_{R}(t) \geq \delta e^{-\alpha(t_{2}-t_{1})}$. We infer that $t_{2}-t_{1} \geq \frac{\log (\frac{\delta}{m})}{\alpha}$. Now suppose, we say $Q_{0} > 0$ such that for $Q \geq Q_{0}$. We have the inequality $|\frac{1}{Q} \int_{t_{0}}^{t+Q} (\Lambda_{R}(t)-\mu_{R}) dt - (\tilde{(\Lambda_{R} - \mu_{R})})| < \delta$, by the uniform convergence hypothesis. Suppose now we say $m < \delta$, such that $t_{2} - t_{1} \geq Q_{0}$ and $(t_{2}-t_{1})(p \tilde{\Lambda_{R}}-\mu_{R}-\delta -W \delta) > 0$, find this to be possible since $(\tilde{\Lambda_{R} - \mu_{R} - \delta - W \delta})>0$. We then have, $\frac{dN_{R}}{dt}\geq (p(t) - WN_{R})N_{R} \geq (p(t) - WN_{R})N_{R}$ for $t_{1} \leq t \leq t_{2}$ such that $m= N_{R}(t) \geq \delta e^{\int^{t_1}-{t_2}(p(t) - W N_{R})N_{R}} dt \geq \delta
e^{(t_1-t_2)}(\tilde{\Lambda_{R}} - \mu_{R} - \delta - W N_{R} \delta) > \delta$ which is contradiction. Therefore $N_{R}(t) \geq m$ and $N_{R}(t)$ is bounded below as asserted.\\
Following Maia Martcheva (2015), we recall that $\frac{dN_{R}}{dt}=\Lambda_{R}(t) N_{R} (1 - \frac{N_{R}}{M})-\mu_{R} N_{R}$
where $N_{R}=S_{R} +I_{R}$. \\
This implies that\\
$\frac{dN_{R}}{dt}\leq (\Lambda_{R}(t) - \mu_{R}) N_{R} - \frac{\Lambda_{R}(t) N^{2}_{R}}{M})$\\
We have shown before that in this case,\\
$\limsup_{t\rightarrow \infty} N_{R}(t) \leq \frac{ M(\Lambda_R - \mu_R)}{\Lambda_R}$\\
Given $\tau > 0$ such that\\
$R_{L_{2}}(\tau)=\frac{(\langle \Lambda_{R}\rangle-\mu_{R})\beta_{R} M}{\langle \Lambda_{R}\rangle \mu_{R} N_{R}} < 1$\\
$\exists$ $t_{0}$ such that $\forall$ $t>t_{0}$, we obtain\\
$N_{R}(t)<\frac{M(\Lambda_{R}(t) - \mu_{R})}{\Lambda_{R}(t)}+ \tau = M(\frac{\Lambda_{R}(t) - \mu_{R}}{\Lambda_{R}(t)}+ \tau)$ is true $\forall$ $t>0$.\\
By considering the equation for $I_{R}$, which is given by\\
$\frac{dI_{R}(t)}{dt}\leq | \beta_{R} M (\frac{\Lambda_{R}(t) - \mu_{R}}{\Lambda_{R}(t)}+ \tau) - \mu_{R}|I_{R}(t)$\\
Solving the inequality above we obtain\\
$I_{R}(t)\leq I_{R}(0) \exp^{\int^{t}_{0}}| \beta_{R} M(\frac{\Lambda_{R}(t) - \mu_{R}}{\Lambda_{R}(t)}+ \tau) - \mu_{R}| dq$\\
Therefore, if $R_{L_{2}}(\tau)<1$, when $I_{R}(t) \rightarrow 0$ as $t \rightarrow \infty$. $\Box$.\\
We assume for simplicity that \\
$S_{R}(0)+I_{R}(0)=\frac{M(\Lambda_{R}(t) - \mu_{R})}{\Lambda_{R}(t)}$\\
Then\\
$N_{R}(t) = \frac{M(\Lambda_{R}(t) - \mu_{R})}{\Lambda_{R}(t)}$ $\forall$ $t$\\
Therefore, we say\\
$S_{R}(t) = \frac{M(\Lambda_{R}(t) - \mu_{R})}{\Lambda_{R}(t)} - I_{R}(t)$\\
From the compartment for the \textit{Mastomys Natalensis} Rat population, we obtain the single one compartment equation in $I_{R}$ given by
\begin{equation}
\begin{array}{lcl}
\ \\
\frac{dI_{R}(t)}{dt} & = & \frac{\beta_{R} M (\Lambda_{R}(t) - \mu_{R})}{\Lambda_{R}(t) N_{R}}(1-I_{R})I_{R} - \mu_{R}I_{R}(t)\\
\ \\
\end{array}
\end{equation}
Using the above notations and concepts we summarised in Lemma 3 below.\\
\textbf{Lemma 3.} The disease free equilibrium (DFE) is globally stable if $I_{R}(t)$ goes to zero as t tends to infinity whenever $R_{L_{2}}<1$.\\
\textbf{Lemma 4.}:If $\Lambda_{R}(t)$ is a periodic function with period $Q$ and let $R_{L_{2}}>1$ then equation $(10)$ has a unique periodic solution $\eta(t)$ (Maia Martcheva, 2015)\\
\textbf{Proof:} First we define the region $J_{R} = \{ I_{R} : I_{R} \in [0,N_{R}]\}$ such that the equation $(10)$ is considered on this region.
We applied here the Poincare map $\mathbb{P}$ to establish the existence of a periodic solution.
It obvious that by the Poincare map we define
$$\mathbb{P}(I_{R}):[0,N_{R}]\rightarrow [0,N_{R}]$$
which implies
$$\mathbb{P}(I_{R}(0))=I_{R}(Q,I_{R}(0))$$
such that $I_{R}(Q,I_{R}(0)$ is the value of solution at time $t=Q$.\\
We understand that the Poincare map is injective because of the properties of solutions of ODEs. Hence, we can further show that it is continuously differentiable. Hence, it is easy to show that $\mathbb{P}(0)=0$ $\&$ $\mathbb{P}(N_{R}< N_{R}$.
Since the number $I_{R} \in [0, N_{R}]$ is an initial value of a periodic solution if and only if $I_{R}$ is a fixed point of the Poincare map. Hence, for one to establish the existence of a positive periodic solution of equation $(10)$, it is expedient to show that the Poincare map has a fixed point. Hence, we define $$h(t)=\frac{\partial I_{R}}{\partial I_{R}(0)}(t,I_{R}(0))$$
We obtain the derivative of the Poincare map which is given as
$$\mathbb{P}'(I_{R}(0)= \frac{\partial I_{R}}{\partial I_{R}(0)}(Q,I_{R}(0))=h(Q)$$
Hence, by differentiating equation $(11)$ with respect to $I_{R}(0)$ in order to obtain the derivative of the Poincare map. For this case, we obtain a differential equation in $h$ such that\\
$\frac{dh(t)}{dt}=h(t)[\frac{\beta_{R} M (\Lambda_{R}(t)- \mu_{R})I_{R}(t,I_{R}(0))}{\Lambda_{R}(t)N_{R}} - \mu_{R}I_{R}(t)]$\\
By differentiating the $I_{R}(0)$ which is the initial condition with respect to $I_{R}(0)$, we have that $h(0)=1$. For the derivative of the Poincare map, which gives the following expression, the differential equation for $h$ can be obtained as follows:
$$\frac{d\mathbb{P}(I_{R}(0))(t)}{dt}=\exp^{\int^{Q}_{0} [\frac{\beta_{R} M (\Lambda_{R}(t)- \mu_{R})I_{R}(t,I_{R}(0))}{\Lambda_{R}(t)N_{R}} - \mu_{R}I_{R}(t,I_{R}(0))]dt}$$
It is clear that $\frac{d\mathbb{P}(I_{R}(0))(t)}{dt}>0$ and its obvious that the Poincare map is increasing.  Given that, if $I_{R_1}$ and $I_{R_2}$ are two initial conditions which satisfies $I_{R_1}<I_{R_2}$, then we obtain\\
 $\frac{d\mathbb{P}(0)}{dt}=\exp^{Q(\frac{\beta_{R}(\langle \Lambda_{R} \rangle- \mu_{R})}{\langle \Lambda_{R} \rangle}\frac{M}{N_{R}} - \mu_{R})}$
This implies that the exponent since $R_{L_2}>1$ and that $\frac{d\mathbb{P}(0)(t)}{dt}>1$. Thus, for $I_{R}(0)$ sufficiently small:
$$\frac{\mathbb{P}(I_{R}(0))-\mathbb{P}(0)}{I_{R}(0)}\simeq\frac{d\mathbb{P}(0)}{dt}>0$$
It means that $I_{R}(0)$ is small enough for $\mathbb{P}(I_{R}(0))> I_{R}(0)$. For $\mathbb{P}(N_{R})<N_{R}$ which implies that the function $\mathbb{P}(I_{R}(0))-I_{R}(0)$ changes sign in the interval $(0,N_{R})$. Therefore, there should exist an $I_{m}$ such that it becomes zero, which means that $$\mathbb{P}(I_{m})=I_{m}$$.
For us to establish the uniqueness of a periodic solution. It is essential to assume that there are two kinds of periodic solutions $I_{m_1}$ $\&$ $I_{m_2}$.
Hence, assume without loss of generality that
$$I_{m_1}<I_{m_2}$$
Let $I_{m}$ be a periodic solution which satisfies firstly the equation $(10)$
\begin{equation}
\begin{array}{lcl}
\ \\
\int^{Q}_{0} |\frac{\beta_{R}(\Lambda_{R}(t)-\mu_{R})}{\Lambda_{R}(t)}(\frac{M}{N_{R}}-I_{R}(t,I_{m}))-\mu_{R}|dq=0
\ \\
\end{array}
\end{equation}
For $I_{m_1}$ and $I_{m_2}$, we obtain
\begin{equation}
\begin{array}{lcl}
\ \\
|I_{m_1}-I_{m_2}=|\mathbb{P}(I_{m_1})-\mathbb{P}(I_{m_2})|=|\frac{d \mathbb{P}(I_{v})}{dt}||I_{m_1}-I_{m_2}|
\ \\
\end{array}
\end{equation}

such that $I_{v}$ satisfies $I_{m_1}<I_{v}<I_{m_2}$, then we have that

\begin{equation}
\begin{array}{lcl}
\ \\
\tiny{\frac{d \mathbb{P}(I_{v})}{dt}=\exp^{\int^{Q}_{0}\frac{\beta_{R}(\Lambda_{R}(t)-\mu_{R})}{\Lambda_{R}(t)}(\frac{M}{N_{R}}-I_{R}(t,I_{m}))
-\mu_{R}-\beta_{R}(\Lambda_{R}(t)-\mu_{R})}{\Lambda_{R}(t)I_{R}(t,I_{m})}<F_{1}dt < F_{2}dt < 1}
\ \\
\end{array}
\end{equation}
Where \\
$F_{1} = exp^{\int^{Q}_{0}\frac{\beta_{R}(\Lambda_{R}(t)-\mu_{R})}{\Lambda_{R}(t)}(\frac{M}{N_{R}}-I_{R}(t,I_{m}))
-\mu_{R}-\beta_{R}(\Lambda_{R}(t)-\mu_{R})}{\Lambda_{R}(t)}I_{R}(t,I_{m})$\\

$F_{2} = exp^{\int^{Q}_{0}\frac{\beta_{R}(\Lambda_{R}(t)-\mu_{R})}{\Lambda_{R}(t)}I_{R}(t,I_{v})}{\Lambda_{R}(t)}I_{R}(t,I_{v})$\\
Therefore, the result showed that the contradiction we obtained in $(12)$ is as a result of the assumption that we have two distinct positive periodic solutions.\\
\textbf{Lemma 5.}:Suppose $\Lambda_{R}(t)$ is a periodic function with the period $Q$ and assume that $\tilde{R}_{L_2}>1$,then the unique periodic solution $\eta(t)$ of eq.(10) is globally stable, that is, if $I_{R}(t,I_{R}(0))$ is a solution with initial condition $I_{R}(0)$ then
\begin{equation}
\begin{array}{lcl}
\ \\
\lim_{t\rightarrow \infty}|I_{R}(t,I_{R}(0)|=\eta(t)\\
\ \\
\end{array}
\end{equation}
\textbf{Proof.}: Here, we first show the convergence to the periodic solution and consider the solutions of eq. $(10)$ by assuming that $R_{L_2}>1$. Suppose $I_{R}(t)$ is an arbitrary solution with the initial condition $I_{R}(0)$.
Recall that $I_{v}$ is the initial condition for the periodic solution. if we assume that $I_{v}=I_{R}(0)$ we have two choices,
\begin{itemize}
  \item $\mathbb{P}(I_{R}(0))> I_{R}(0)$
  \item $\mathbb{P}(I_{R}(0))< I_{R}(0)$
\end{itemize}
Lets assume that $\mathbb{P}(I_{R}(0))< I_{R}(0)$ and then the second option can be taken care of in the same manner. We have $$\mathbb{P}^{k}(I_{R}(0))< \mathfrak{P}^{k-1}(I_{R}(0))$$.
Therefore, the sequence $\mathbb{P}^{k}(I_{R}(0))$ is a decreasing sequence. Hence, it must converge to a limit since it is bounded below
$$\lim_{k\rightarrow \infty} \mathbb{P}^{k}(I_{R}(0)) = I_{R}(\infty)$$
The number $I_{R}(\infty)$ is the limit of the sequence which is a fixed point of the Poincare map $ \mathbb{P}(I_{R}(\infty)) = I_{R}(\infty)$. The Poincare map of model equation $(11)$ has only two fixed points $I_{R}(\infty)=I_{v}$.\\
If $I_{R}(\infty)=0$ then for some $N_{R}$, the number $\mathbb{P}^{N_{R}}(I_{R}(0))$ is small enough that from the properties of the Poincare map, $$\mathbb{P}^{N_{R}+1}(I_{R}(0))>\mathbb{P}^{k}(I_{R}(0))$$ which contradicts the fact that the sequence is decreasing . Hence, $I_{R}(\infty)=I_{v}$ as a result, the limit in (14) holds.
\section{Existence of Endemic Equilibrium Point}
Here, we present the existence of the endemic equilibrium point $J_{EE}$ for the Lassa fever periodically-forced model $(1)$. It is a non-negative equilibrium state where the Lassa fever disease persists in the population.\\
\textbf{Theorem 4.}:Suppose there exists a unique endemic equilibrium point when $R_{0}>1$ in the Lassa fever periodically forced model $(1)$.\\
\textbf{Proof.}: Suppose $J_{EE}=(S^{\ast \ast}_{H},E^{\ast \ast}_{H},A^{\ast \ast}_{H}, I^{\ast \ast}_{H}, R^{\ast \ast}_{H},S^{\ast \ast}_{R},I^{\ast \ast}_{R})$ is a non-trivial equilibrium of the model equation $(1)$, which implies that all components of $J_{EE}$ are positive.Setting the LHS of the model equation (1) to zero we obtained the following;

$$S^{\ast \ast}_{H}=\frac{(\Lambda_{H} + \mu_{H}) \Lambda_{H} + \sigma \gamma_{2} I^{\ast \ast}_{H} + \sigma \gamma_{1} A^{\ast \ast}_{H} }{(\Lambda_{H} + \mu_{H}) \mu_{H}(\beta_{H}  I^{\ast \ast}_{R} + \beta_{1} A^{\ast \ast}_{H} + \beta_{2}  I^{\ast \ast}_{H})}$$

$$E^{\ast \ast}_{H}=\frac{(\gamma \gamma_{2} I^{\ast \ast}_{H} +  \Lambda_{H} (\theta + \mu_{H}))[\beta_{3}(\omega_{1} + \gamma_{1} + \mu_{H})(N_{R} M \Lambda_{R}(t) \beta_{R} - M \mu^{2}_{R} - \Lambda_{R}(t) N^{2}_{R} \beta_{R})]}{\beta_{3}(\omega_{1} + \gamma_{1} + \mu_{H})(N_{R} M \Lambda_{R}(t) \beta_{R} - M \mu^{2}_{R} - \Lambda_{R}(t) N^{2}_{R} \beta_{R})(\theta + \mu_{H}) (\theta + \mu_{H}) - \sigma \gamma_{1} (1-\upsilon)\theta N_{H}}$$

\small{$A^{\ast \ast}_{H}=\frac{(1-\upsilon)(\sigma \gamma_{2} I^{\ast \ast}_{H} +  \Lambda_{H} (\theta + \mu_{H}))[\beta_{3}(\omega_{1} + \gamma_{1} + \mu_{H})(N_{R} M \Lambda_{R}(t) \beta_{R} - M \mu^{2}_{R} - \Lambda_{R}(t) N^{2}_{R} \beta_{R})]}{\beta_{3}((\omega_{1} + \gamma_{1} + \mu_{H})(N_{R} M \Lambda_{R}(t) \beta_{R} - M \mu^{2}_{R} - \Lambda_{R}(t) N^{2}_{R} \beta_{R})(\theta + \mu_{H}) (\theta + \mu_{H}) - \sigma \gamma_{1} (1-\upsilon)\theta)(\omega_{1} + \gamma_{1} + \mu_{H})(N_{R} M \Lambda_{R}(t) \beta_{R} - M \mu^{2}_{R} - \Lambda_{R}(t) N^{2}_{R} \beta_{R})}$}

$I^{\ast \ast}_{H}=\frac{\upsilon \theta (\gamma_{2}I^{\ast \ast}_{H} + \Lambda_{H} (\theta + \mu_{H})) ((\omega_{1} + \gamma_{1} + \mu_{H})\beta_{3} (N_{R} M \Lambda_{R}(t) \beta_{R} - M \mu^{2}_{R} - \Lambda_{R}(t) N^{2}_{R} \beta_{R}))}{\beta_{3}(\omega_{1} + \gamma_{1} + \mu_{H})(N_{R} M \Lambda_{R}(t) \beta_{R} - M \mu^{2}_{R} - \Lambda_{R}(t) N^{2}_{R} \beta_{R})(\theta + \mu_{H}) (\theta + \mu_{H}) - \sigma \gamma_{1} (1-\upsilon)\theta N_{H}}$

$\tiny{R^{\ast \ast}_{H}=\frac{Z_{1}}{(\sigma + \mu_{H})(N_{R} \beta_{3}(\omega_{1} + \gamma_{1} + \mu_{H})(M \Lambda_{R}(t) \beta_{R} - M \mu^{2}_{R} - \Lambda_{R}(t) N_{R} \beta_{R})-\sigma \gamma_{1} M \beta_{R} \mu_{R} )}}$

where \\
$Z_{1} = I^{\ast \ast}_{H}\gamma_{2}N_{R}\beta_{3}(\omega_{1} + \gamma_{1} + \mu_{H})(M \Lambda_{R}(t) \beta_{R} - M \mu^{2}_{R} - \Lambda_{R}(t) N_{R} \beta_{R})-\sigma \gamma_{2}\gamma_{1} M  \beta_{R} \mu_{R}I^{\ast \ast}_{H} + \gamma_{1}(\sigma+\mu_{H})(1-\upsilon)\theta \Lambda_{H} M \beta_{R}\mu_{H}+\sigma \gamma_{1}\gamma_{2} I^{\ast \ast}_{H} M \beta_{R} \mu_{R}$

$$S^{\ast \ast}_{R}=\frac{N_R \mu_{R}}{\beta_{R}}$$

$$I^{\ast \ast}_{R}=\frac{N_{R} M \Lambda_{R}(t) \beta_{R} - M \mu^{2}_{R} - \Lambda_{R}(t) N^{2}_{R} \beta_{R}}{M \beta_{R} \mu_{R}}$$

and

$I^{\ast \ast}_{H}$ is a positive solution of an equation which is given as:

\begin{equation}
\begin{array}{lcl}
\ \\
(Q_1 - \beta_{3}\upsilon \theta \gamma_{2} Q_{2})I^{\ast \ast}_{H} -  \beta_{3}\upsilon \theta  \Lambda_{H}Q_2 = 0\\
\ \\
\end{array}
\end{equation}
where \\
$Q_{1} = \beta_{3}(\omega_{1} + \gamma_{1} + \mu_{H})(N_{R} M \Lambda_{R}(t) \beta_{R} - M \mu^{2}_{R} - \Lambda_{R}(t) N^{2}_{R} \beta_{R})(\theta + \mu_{H}) (\theta + \mu_{H}) - \sigma \gamma_{1} (1-\upsilon)\theta N_{H}$\\
$Q_{2} = (\theta + \mu_{H})) ((\omega_{1} + \gamma_{1} + \mu_{H})\beta_{3} (N_{R} M \Lambda_{R}(t) \beta_{R} - M \mu^{2}_{R} - \Lambda_{R}(t) N^{2}_{R} \beta_{R}))$.\\
The positive solution to the equation (15) depends on \\
$(Q_1 - \beta_{3}\upsilon \theta \gamma_{2} Q_{2})$ and $\beta_{3}\upsilon \theta  \Lambda_{H}Q_2$ respectively. Hence, there exist a unique endemic equilibrium given by\\
$I^{\ast \ast}_{H} = \frac{\beta_{3}\upsilon \theta  \Lambda_{H}Q_2}{Q_1 - \beta_{3}\upsilon \theta \gamma_{2} Q_{2}}$ and $\beta_{3}\upsilon \theta  \Lambda_{H}Q_2 < 0$, whenever $\widetilde{R}_{L}>1$.
\section{Evaluation of Lassa fever Intervention strategies}
The intervention strategies available for Lassa fever disease are capable of reducing the mortality of humans due to the disease if applied early and reducing the morbidity of Lassa fever in individual if preventive intervention strategies are taken. The following are the various interventions available to reduce the mortality pr morbidity of Lassa fever disease:
\begin{itemize}
  \item Early Treatment with Ribavirin
  \item Community Hygiene to discourage Rodents from entering homes
  \item Isolation of infected humans suspected with symptoms of Lassa fever
  \item Culling/destruction of Rodents with pesticides
  \item Educational Campaign
\end{itemize}
%and tackling the source of infection of the isolated individuals
Here, we create scenarios for our intervention strategies. We understand that these multiple intervention packages are being applied differently in different nations. We need the collection of the opinions of various experts from different disciplines which include, modellers, Public Health officers, entomologists, epidemiologists, etc to decide along with the local national control programs on Lassa fever disease that can bring the disease into elimination locally. The scenario we are building here states that we minimize $I_{H}$ (the number infected humans), which is the number of cases of Lassa fever in humans. We know that each of the intervention packages impacts some of the parameter values. We hereby evaluate the change that a $1\%$ change in a parameter $q$ makes on $I^{\ast}_{H}$ through the principle of elasticity.The elasticity is the percentage change in a static quantity (e.g. Reproduction number, equilibrial prevalence, or equilibrial incidence) with respect to the percentage change in any given parameter in the model. Therefore, the elasticity of any static quantity with respect to any given parameter is positive if the static quantity is increasing with respect to any given parameter and negative if the static quantity is decreasing with respect to any given parameter. Hence, we say that the elasticity of $I^{\ast}_{H}$ concerning the parameter $q$ is given by\\
$\gamma^{q}_{I^{\ast}_{H}} = \frac{\partial I^{\ast}_{H}}{\partial q}\cdot \frac{q}{I^{\ast}_{H}} \approx \frac{\% \Delta I^{\ast}_{H}}{\% \Delta q}$ \\
where\\
$I^{\ast}_{H}=\frac{(\upsilon \theta N_{H} M \beta_{R} \mu_{R}  +  (1-\upsilon) \theta N_{H} M \beta_{R} \mu_{R}) \Lambda_{H}  B_{1}(\theta + \mu_{H}) }{Q}$\\
such that\\
$A_{1}=(\omega_{1} + \gamma_{1}+\mu_{H})(\omega_{2} + \gamma_{2}+\mu_{H})(M \Lambda_{R}(t)\beta_{R}N_{R}-M \mu^{2}_{R}-\Lambda_{R}(t)N^{2}_{R}\beta_{R}) N_{H}\beta_{3}$,\\
$B_{1}=\beta_{3}(\omega_{1} + \gamma_{1}+\mu_{H})(M \Lambda_{R}(t)\beta_{R}N_{R}-M \mu^{2}_{R}-\Lambda_{R}(t)N^{2}_{R}\beta_{R})$,\\ and
$Q=A_{1}B_{1}(\sigma + \mu_{H})(\theta + \mu_{H})-\sigma  \gamma_{2}\upsilon\theta N_{H} M \beta_{R} \mu_{R}(\omega_{1} + \gamma_{1}+\mu_{H})-(1-\upsilon) \theta N_{H} M \beta_{R} \mu_{R}\sigma\gamma_{2}-\sigma\gamma_{1}(1-\upsilon) \theta N_{H} M \beta_{R} \mu_{R}$\\

Maple software was used in computing the elasticities with the above evaluated parameters. We present the elasticities in the Table 3 below:
\begin{table}[H]
\caption{\small Table showing elasticities of $I^{\ast}_{H}$}
\centering
\small
\begin{tabular}{lcl}\hline
Parameter&Elasticity\\ \hline
$\Lambda_{H}$&+1\\
$\Lambda_{0}$&-1\\
$\Lambda_{1}$&-0.4937\\
$\phi$&0.0193\\
$\beta_{3}$&-0.988\\
$\beta_{R}$&-0.0000000065347\\
$\sigma$&-0.91700443\\
$\gamma_{1}$&-0.78814\\
$\gamma_{2}$&-0.88662\\
$\mu_H$&-0.2417115\\
$\mu_R$&+1\\
$\theta_{H}$&+1\\
$\omega_{1}$&-0.1364\\
$\omega_{2}$&-0.0301115\\
$M$&-0.000001292\\
$\upsilon$&-0.00000000065976\\\hline
\end{tabular}
\label{tabmx2}
\end{table}
We observed directly from table 3 above that the intervention strategies, like the early treatment of Lassa fever with Ribavirin, has more effectiveness in influencing the prevalence of Lassa fever in humans than others. We determine which parameters each intervention would affect to compare the available intervention strategies. For example, culling affects $\mu_{R}$, $\omega_{1}$, $\omega_{2}$, and $\beta_{R}$, early treatment with Ribavirin affects $\gamma_{1}$, $\gamma_{2}$, $\sigma$, $\omega_{1}$, $\omega_{2}$, and $\beta_{3}$, community hygiene affects $\beta_{3}$, educational campaigns affects $\beta_{3}$,$\omega_{1}$, and $\omega_{2}$ while isolation of infective humans affects $\omega_{1}$ and $\omega_{2}$, Hence, the total effect of intervention strategies is define as the sum of efficacies of the effect of the strategy on each affected parameter. For example, treatment with Ribavirin increases $\gamma_{1}$, $\gamma_{2}$ but decreases $\sigma$, $\omega_{1}$, $\omega_{2}$, and $\beta_{3}$. Therefore, the total efficacy is -0.78814-0.88662-0.91700443-0.1364-0.0301115-0.988=-3.74627593, we take this number as an absolute value. The total effects of each intervention strategy are summarized in Table 4.

\begin{table}[H]
\caption{\small Table showing the list of intervention strategies and their efficacy}
\centering
\small
\begin{tabular}{lccccl}\hline
Intervention&affected parameter&Total efficacy (dimensionless)&Rank \\ \hline
Culling without repopulation&$\mu_{R}$,$\omega_{1}$,$\omega_{2}$, $\beta_{R}$ &0.8334885&4\\
Early Treatment with Ribavirin&$\gamma_{1}$, $\gamma_{2}$,$\sigma$,$\omega_{1}$,$\omega_{2}$, $\beta_{3}$&3.74627593&1\\
Community hygiene&$\beta_{3}$&0.988&3\\
Isolation of infected humans&$\omega_{1}$,$\omega_{2}$&0.1665115&3\\
Educational campaign&$\beta_{3}$, $\omega_{1}$,$\omega_{2}$&1.1545115&2 \\\hline
\end{tabular}
\label{tabmx3}
\end{table}
Table 4 suggested that treatment with ribavirin is the most effective strategy but where resources are not sufficient educational campaign and community hygiene are the next two most efficient and effective intervention strategies followed by culling and lastly isolation of infected cases.\\
\textbf{Remark 5}: The total efficacy is the overall sum of the protection/treatment provided by each intervention strategy that was employed to control and/or eliminate a disease or an outbreak, which excludes the herd effect. The total efficacy is dimensionless.\\

\section{Numerical results and discussion}
In this section, we present the numerical simulation results of our model and its analysis to establish our theoretical findings. The parameters presented in Table 5 below which some were from kinds of literature and some were from assumptions. We simulated the model system (1) using MATLAB ODE45 solvers and the following initial conditions were considered; $S_{H} = 996$, $E_{H}=0$, $A_{H}=4$, $I_{H}=0$, $R_{H}=0$, $S_{R}=5000$, $I_{R}=10$.

\begin{table}[H]
\caption{\small Table showing values and ranges for parameters used in the Lassa fever model}
\centering
\small
\begin{tabular}{lccccl}\hline
Symbol&Dimension&Range&Source \\ \hline
$\Lambda_{H}$&$humans \times day^{-1}$&$1000*0.0003465$& Nakul et al. (2008)\\
$\Lambda_{0}$&$day^{-1}$&$0.033$&Assumed\\
$\Lambda_{1}$&Dimensionless&1.0&Keeling and Rohani,(2008)\\
$\phi$&$day^{-1}$&10&Keeling and Rohani, (2008)\\
$\beta_{H}$&Dimensionless&0.024-0.048&Nakul et al. (2008)\\
$\beta_{1}$&Dimensionless&0.022-0.27&Nakul et al. (2008)\\
$\beta_{2}$&Dimensionless&0.021-0.8&Mohammed et al. (2015)\\
$\beta_{2}$&Dimensionless&0.021-0.8&Mohammed et al. (2015)\\
$\beta_{3}$&Dimensionless&0.24-0.8&Mohammed et al. (2015)\\
$\sigma$&$day^{-1}$&0.00385-&Mohammed et al. (2015)\\
$\gamma_{1}$&$day^{-1}$&0.3333-0.8&Mohammed et al. (2015)\\
$\gamma_{2}$&$day^{-1}$&0.6086-0.8&Mohammed et al. (2015)\\
$\mu_H$&$day^{-1}$&0.0003465& WHO(2016)\\
$\mu_R$&$day^{-1}$&0.00641026-0.0038&Mohammed et al. (2015)\\
$\theta_{H}$&$day^{-1}$&0.3333&CDC\\
$\omega_{1}$&$day^{-1}$&0.00019231&WHO(2016)\\
$\omega_{2}$&$day^{-1}$&0.00019231&WHO (2016)\\
$M$&$day^{-1}$&$3.9\times10^{7}$&Nakul et al. (2008)\\
$\upsilon$&$day^{-1}$&1.5&Assumed \\\hline
\end{tabular}
\label{tabmx1}
\end{table}

\subsection{Time series solution of the periodically-forced Lassa fever model}
It was observed from the first graph in fig. 2a that as the infected humans population increases at an instant of time, which implies that as people get exposed within less than 30days they become infected and infectious but in the long run as the infected humans population decreases until they reached a steady-state, which denotes that the Lassa fever is endemic in the population, even in the absence of the exposed and the asymptomatic humans. It was observed from  the population of the susceptible humans deplete as people get infected with Lassa fever over a long period. This showed that if continuous and consistent interventions are not put in place, a large number of people will be infected in the long run. It was also observed from fig. 2b that the number of susceptible rats/rodents depleted over time as the number of infected rats increases. There was seasonality in the increase rate of infected rats due to seasonality in the birth rate. This depicts that the increase in the number of infected rats also caused an increase in the number of infected humans when there is a successful interaction between humans and infected rats. In the fourth graph under fig. 3, we observed that when people recovered due to treatment with ribavirin the number of recovered humans increases in the population. Treatment with ribavirin alone does not stop the transmission of the disease, it will only reduce mortality but not morbidity. So, we must employ other intervention strategies to eliminate the disease from the community.

\begin{figure}[ht]
\centering
\subfigure[ ]{
   \includegraphics[scale =0.5] {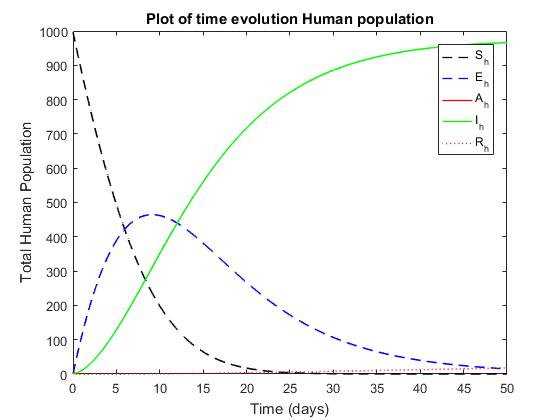}
   \label{fig:subfig1}
}
\subfigure[]{
   \includegraphics[scale =0.5] {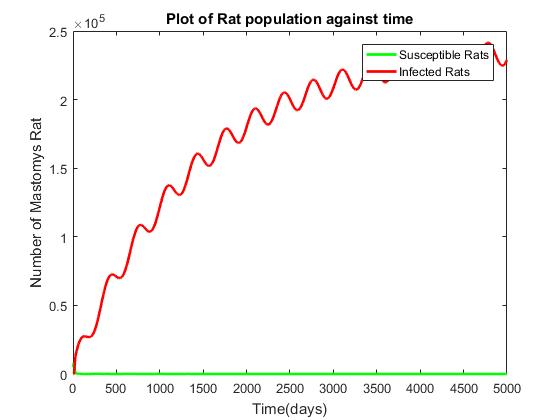}
   \label{fig:subfig2}
}
\label{myfigure}
\small{\caption{Solution of the Lassa fever periodically-forced model (1) with parameter values defined in Table 5. The Initial condition is $S_{H} = 996$, $E_{H}=0$, $A_{H}=4$, $I_{H}=0$, $R_{H}=0$, $S_{R}=5000$, $I_{R}=10$.}}
\end{figure}

\subsection{Behaviour of the time dependent per capita birth rate of rodents with varying values of $\Lambda_{1}$}
It was observed in fig. 3 below that as we vary the value of the degree of seasonality ($\Lambda_{1}$), it increases the peak, the magnitude, and the length of the seasonality of the time-dependent per capita birth rate of the\textit{ mastomys} rat. The higher the degree of seasonality, the higher the effect of seasonality.
\begin{figure}[ht]
\centering
\includegraphics[trim = 0mm 0mm 0mm 0mm,clip,width=10cm]{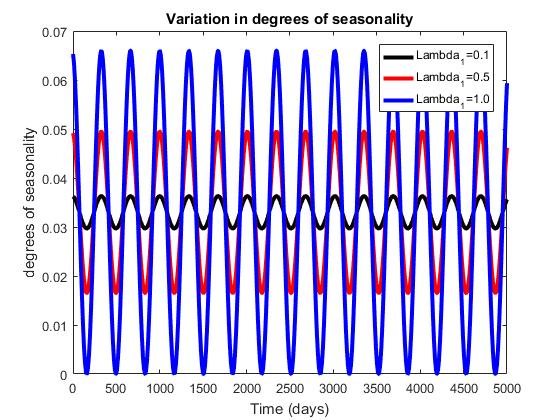}
\caption{Behaviour of the time-dependent per capita birth rate of \textit{mastomys natalensis} rat as it varies with different values of $\lambda_{1}$ such that $\Lambda_{1}=0.1$,$\Lambda_{1}=0.5$,$\Lambda_{1}=1.0$.}
\end{figure}
\newpage
\subsection{Behaviour of the Infected\textit{mastomys} rodents with varying values of $\Lambda_{1}$}
It was observed in fig. 4 below that as we increase the value of the degree of seasonality ($\Lambda_{1}$), the peak, the magnitude, and the length of the seasonality of the infected \textit{mastomys} rat population increases over a long period.
\begin{figure}[ht]
\centering
\includegraphics[trim = 0mm 0mm 0mm 0mm,clip,width=10cm]{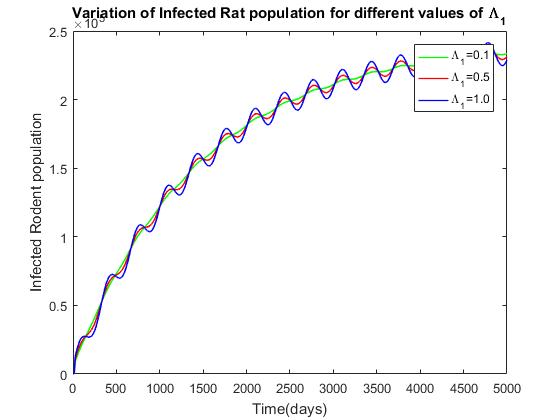}
\caption{Variation in the degrees of seasonality, we varied different values of $\Lambda_{1}$ such that $\Lambda_1=0.1$,$\Lambda_1=0.5$,$\Lambda_1=1.0$.}
\end{figure}

\subsection{Behaviour of the Infected\textit{mastomys} rodent against Infected humans}
It was observed in fig. 5 below that as the number of Infected \textit{mastomys} rodent increases, it implies increase in Lassa fever disease prevalence.This depicts that the number of Infected \textit{mastomys} rodent has a great role to play in humans Lassa fever disease prevalence.
\begin{figure}[ht]
\centering
\includegraphics[trim = 0mm 0mm 0mm 0mm,clip,width=10cm]{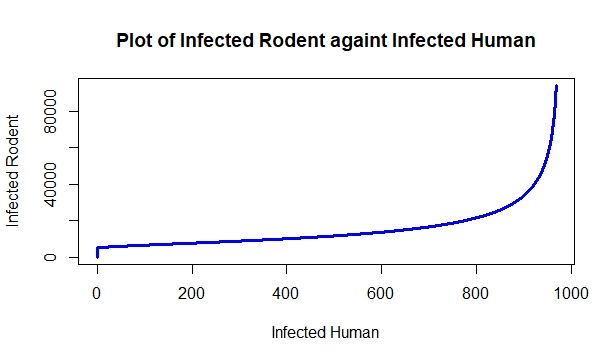}
\caption{Phase plane for Infected \textit{mastomys} rodent against Infected humans.}
\end{figure}
\newpage
\section{Conclusion, Recommendation and Future work}
Based on the works of Bawa et al. (2014), where they modelled the dynamics of Lassa fever by subdividing the rodent population into infant and adult classes and carried out analyses on their model to establish global stability condition for the control of Lassa fever, Mohammed et al. (2015), where they developed a transmission dynamic model for Lassa fever with humans immigration. Model analyses were carried out to calculate the reproduction number and sensitivity analysis of the model was also performed. Their results showed that humans immigration rate is the most sensitive parameter and then the humans recovery rate is the second most sensitive parameter followed by person-to-person contact rate. Hence, they suggested that control strategies should target humans immigration, effective drugs for treatment and education to reduce person-to-person contact. Andrei et al. (2019), where they had a simulation based model and showed from their work that seasonal migratory dynamics of rodents played a key role in regulating the cyclic pattern of Lassa fever epidemics and Joachim et al. (2019) also developed a model that suggested the use of continuous control or rodent vaccination are the strategies that can lead to Lassa fever elimination. All these works did not considered the impact of periodically-forced per-capita birth rate on the transmission dynamics of Lassa fever. Hence, in this work, we presented a periodically-forced Lassa fever model to understand the transmission dynamics of Lassa fever disease in the host population. It was observed that the yearly average is given by $\widetilde{R}_{L}<1$, when the disease does not invade the population (means that the number of infected humans always decrease in the following seasons of transmission) and $\widetilde{R}_{L}>1$ when the disease remains constant and is invading the population and it was detected that $\widetilde{R}_{L} \neq R_{L}$. Rigorous mathematical analyses of the model were carried out to study the qualitative properties of the model through which existence and uniqueness of the periodic solution with its global stability were established.Existence of the endemic equilibrium point for the model was also explored and a unique endemic equilibrium exist whenever $\widetilde{R}_{L}>1$. Analysis of the Lassa fever intervention strategies revealed that the use of multiple intervention strategies must be employed to mitigate against Lassa fever disease by reducing its morbidity and mortality. It was established that to eliminate Lassa fever disease, treatments with Ribavirin must be provided early to reduce mortality and other preventive measures like the educational campaign, community hygiene, Isolation of infected humans, and culling/destruction of rodents with pesticides must be applied to also reduce the morbidity of the disease. There are still many questions to answer in the modelling of Lassa fever transmission dynamics. It will be our interest to have access to country-specific observational or serological data that we can use to validate our model. We are particularly interested in the data from Nigeria were we have had a lot of epidemic of Lassa fever in recent times. We also intend to apply other techniques like optimal intervention analysis that will help us to set up some intervention scenarios and guide us in choosing the best strategies, stochastic simulations will also be of interest, which will help us capture the random nature of the disease in the community. We strongly believe there are many works to be done in the modelling and epidemiology of Lassa fever disease. Finally, the results obtained in this work can be a useful guide for the local national control program on Lassa fever, decide on the framework for planning and designing the cost-effective strategies for the best intervention packages in eliminating Lassa fever in Nigeria and West Africa at Large.

\section{Acknowledgement}
All the authors would like to thank TETFUND Institution Based Research Fund (IBRF) Ref No.: TETF/DAST and D.D./6.13/NOMCA/BAS (P.V. No. 090403) on Lassa Fever Modelling Project,Federal Government of Nigeria for the financial support to carry out the present work.

\end{document}